\documentclass{article}

\usepackage{arxiv}

\usepackage[utf8]{inputenc} 
\usepackage[T1]{fontenc}    
\usepackage{hyperref}       
\usepackage{url}            
\usepackage{booktabs}       
\usepackage{amsfonts}       
\usepackage{nicefrac}       
\usepackage{microtype}      
\usepackage{lipsum}
\usepackage{graphicx}
\usepackage{amsmath}
\usepackage{breqn}
\graphicspath{ {./images/} }
\usepackage{natbib}

\title{Mitigating cycle skipping in full waveform inversion using max-pooling-based approximate envelope and shot patching}

\author{
 Xinru Mu \\
  Physical Science and Engineering Division\\
  King Abdullah University of Science and Technology\\
  Thuwal 23955, Saudi Arabia \\
   \And
Omar M. Saad \\
  Physical Science and Engineering Division\\
  King Abdullah University of Science and Technology\\
  Thuwal 23955, Saudi Arabia \\
  \And
Shaowen Wang \\
  Physical Science and Engineering Division\\
  King Abdullah University of Science and Technology\\
  Thuwal 23955, Saudi Arabia \\
  \And  
 Tariq Alkhalifah \\
  Physical Science and Engineering Division\\
  King Abdullah University of Science and Technology\\
  Thuwal 23955, Saudi Arabia \\
}

\begin{document}
\maketitle

\begin{abstract}
Full waveform inversion (FWI) can produce accurate subsurface velocity models. However, the lack of sufficiently low-frequency content in field data often causes cycle skipping and traps the inversion in local minima. The Hilbert-transform envelope (HTE) provides a low-frequency representation that helps mitigate cycle skipping, but it may be insufficient when the initial velocity model is highly inaccurate. To further enhance low-frequency information and reduce dependence on the initial model, we compute an approximate envelope using a sequence of 2D max-pooling operations. Compared with HTE, the resulting max-pooling-based approximate envelope (MPBAE) contains richer low-frequency components and better mitigates cycle skipping. We further combine the MPBAE loss with a shot patching strategy and exploit the inherent normalization property of the Euclidean loss to formulate the MPBAEP loss, in which each shot gather is divided into localized patches for misfit evaluation. This introduces local adjoint-source energy balancing, as the adjoint source associated with the Euclidean loss exhibits a normalization effect within each local region, thereby improving gradient balance and accelerating convergence. Numerical experiments on synthetic and field data demonstrate that MPBAE-FWI significantly outperforms HTE-FWI when the initial model is poor, while MPBAEP-FWI further improves inversion accuracy.
\end{abstract}


\section{Introduction}
Full waveform inversion (FWI) is a high-resolution method for constructing subsurface velocity models by iteratively updating the model to reduce the discrepancy between observed and simulated data \citep{tarantola1984inversion}. The objective function in FWI plays a crucial role in obtaining accurate inversion results. Ideally, a convex objective function ensures that FWI converges to the global minimum corresponding to the true subsurface model. However, in practice, the FWI objective function is often highly non-convex due to the band-limited nature of seismic data and complex nature of wave phenomena. This non-convexity can lead the inversion to converge to a local minimum, making it highly dependent on the initial model and the low-frequency content in the data \citep{virieux2009overview, oh2018full}.

Designing strategies to improve the convexity of the misfit function is critical for achieving robust and accurate velocity inversion results. The optimal transport (OT) objective function has demonstrated strong potential in mitigating cycle-skipping in FWI by quantifying the discrepancy between observed and simulated seismic data through the minimal cost of transporting one distribution into another, thereby providing a physically meaningful assessment of both amplitude and phase differences \citep{engquist2016optimal, sun2019application, da2022graph}. Another effective strategy involves constructing misfit functions using matching filters derived from the deconvolution of observed and simulated traces \citep {warner2016adaptive, sun2020joint, yong2023localized}. In this framework, optimization is achieved by penalizing the filter coefficients at non-zero time lags, as the ideal filter reduces to a band-limited Dirac delta function when the simulated data perfectly match the observations. Additionally, by leveraging the intrinsic ultra-low-frequency information in seismic data, the Hilbert-transform-envelope-based (HTE-based) loss function can mitigate cycle skipping and provide improved initial models for FWI \citep{wu2014seismic, oh2018full, chen2022salt, song2023weighted}. However, HTE-based loss functions still struggle to produce a reliable inversion result when the starting velocity model is severely inaccurate. In recent years, several machine learning–based loss functions, such as FWIGAN\citep{yang2023fwigan} and SiameseFWI \citep{saad2024siamesefwi}, have also been introduced to improve inversion accuracy.

Data patching is a common strategy in machine learning, mainly used to decrease memory usage and enable efficient handling of large datasets \citep{brutzkus2022efficient}. By splitting the input data into smaller, more manageable segments, it not only allows for training on high-resolution data that would otherwise exceed GPU memory, but also facilitates learning of local features that can enhance model performance and generalization \citep{hett2020patch}. In light of this, we leverage the capability of patching for local data comparison and incorporate it into the construction of the FWI loss function. Patching methods can generally be classified into overlapping and non-overlapping types. In overlapping patching, neighboring patches share part of their regions, which helps reduce boundary artifacts and maintain continuity in predictions. Non-overlapping patching, on the other hand, is simpler and more memory-efficient, though it may necessitate additional post-processing to address edge effects \citep{pielawski2020hann}. 

In this study, we obtain an approximate seismic envelope by applying a series of 2D max-pooling operations. Compared with the conventional HTE, the max-pooling-based approximate envelope (MPBAE) admits stronger low-frequency components, which can further reduce the dependence of FWI on the initial velocity model, while maintaining the high-resolution properties of conventional FWI. We construct a new envelope-based objective function using MPBAE and refer to the resulting inversion scheme as MPBAE-FWI, which is benchmarked against the traditional HTE loss, with its corresponding implementation denoted as HTE-FWI. In addition, we combine non-overlapping shot patching with both the Euclidean loss and the MPBAE loss, resulting in the proposed MPBAEP loss. In this framework, each shot gather is divided into patches, and the adjoint source associated with the Euclidean loss inherently normalizes the energy within each patch, thereby generating gradients with more balanced amplitudes. Convexity analysis of the objective function indicates that the MPBAE loss is more convex than the HTE loss, thereby mitigating cycle skipping. Furthermore, the MPBAEP loss also exhibits better convexity than the MPBAE loss. Numerical experiments on synthetic and field datasets demonstrate that MPBAE-FWI significantly outperforms HTE-FWI in recovering subsurface velocity when the initial model is highly inaccurate. In addition, MPBAEP-FWI further improves inversion accuracy compared with MPBAE-FWI.

\section{Theory}
In this section, we first review the fundamental theory of FWI, followed by an introduction to the MPBAE and MPBAEP loss functions, along with an analysis of the mechanisms by which they mitigate cycle skipping and improve inversion accuracy.

\subsection{Fundamentals of acoustic FWI}
The acoustic wave equation in the time domain in 2D space can be written as: 
\begin{equation}
\label{eq1}
\frac{{{\partial ^2}u(\textbf{x},t)}}{{\partial {t^2}}} = {v^2}({\bf{x}})\left( {\frac{{{\partial ^2}u(\textbf{x},t)}}{{\partial {x^2}}} + \frac{{{\partial ^2}u(\textbf{x},t)}}{{\partial {z^2}}}} \right) + f({\textbf{x}_s},t)\delta (\textbf{x} - {\textbf{x}_s}),
\end{equation}
where \(u\) is the pressure wavefield as a function of spatial location \({\bf{x}} = (x,{\rm{ }}z)\) and time \(t\), \(v\) denotes the seismic wave propagation velocity, and \(f({\textbf{x}_s},t)\) is the source time function located at the source location \({\textbf{x}_s}\). The Dirac delta function \(\delta \) is used to model a point source at \({\textbf{x}_s}\).

FWI relies on an objective function to measure the mismatch between observed and simulated data. Here, we adopt the Euclidean loss, which is mathematically defined as follows:

\begin{equation}
\label{eq2}
{L_{{Eucl}}} = \sqrt {\sum\limits_{s,{\rm{ }}r,{\rm{ }}t} {{{\left( {{d_{sim}}(s,r,t) - {d_{obs}}(s,r,t)} \right)}^2}} }, 
\end{equation}
where \(s\), \(r\), and \(t\) represent the source, receiver, and time indices, respectively, \({{d_{sim}}}\) and \({{d_{obs}}}\) are the simulated and observed data. By comparison with the L2 loss, the Euclidean loss is defined as the square root of the L2 loss.

FWI typically updates the velocity model \({\bf{m}}\) iteratively using the gradient descent method, which can be written as:
\begin{equation}
\label{eq3}
{{\bf{m}}_{k + 1}} = {{\bf{m}}_k} - \lambda \frac{{\partial L}}{{\partial {{\bf{m}}_k}}},
\end{equation}
where \(L\) is the loss function, \(\partial L/\partial {{\bf{m}}_k}\) denotes the gradient of the objective function with respect to the velocity model at the \textit{k}-th iteration, and \(\lambda \) represents the step length used for velocity updating.

\subsection{Max-pooling and MPBAE loss}
If equation (\ref{eq2}) is used as the FWI loss function, the inversion becomes highly prone to cycle skipping when the observed data lack low-frequency content and the initial velocity model is poor. To mitigate this issue, we can use the envelope of the seismic data, which is extracted using the Hilbert transform. The HTE-based objective function can be defined as follows:

\begin{equation}\label{eq4}
{L_{HTE}} = \sqrt {\sum\limits_{s,{\rm{ }}r,{\rm{ }}t} {{{\left( {e_{sim}^p(s,r,t) - e_{obs}^p(s,r,t)} \right)}^2}} },
\end{equation}
where \({e_{sim}} = \sqrt {d_{sim}^2 + d_{Hsim}^2} ,\)~\({e_{obs}} = \sqrt {d_{obs}^2 + d_{Hobs}^2} ,\) and \({d_{Hsim}}\) and \({d_{Hobs}}\) are the Hilbert-transformed counterparts of \({d_{sim}}\) and \({d_{obs}}\). Here, \(p\) is the envelope exponent, which can take any positive value.

To further enhance the low-frequency content of the envelope and reduce FWI’s sensitivity to the initial velocity model, we compute an approximate envelope by applying a sequence of max-pooling operations to the data. We define \(q\) as the number of max-pooling operations. The objective function based on the MPBAE can be written as follows:
\begin{equation}\label{eq5}
{L_{MPBAE}} = \sqrt {\sum\limits_{s,{\rm{ }}r,{\rm{ }}t} {{{\left( {{\zeta _{sim}}(s,r,t) - {\zeta _{obs}}(s,r,t)} \right)}^2}} },
\end{equation}
where \({{\zeta _{sim}}}\) and \({{\zeta _{obs}}}\) are the MPBAEs obtained by applying max pooling to \({d_{sim}}\) and \({d_{obs}}\), respectively. The max-pooling operation is defined as:
\begin{equation}\label{eq6}
{Y_{i,j}} = \mathop {\max }\limits_{0 \le a < {k_{h,}}{\rm{ }}0 \le b < {k_w}} {X_{{\kern 1pt} i{s_h} + a,\;j{s_w} + b}},\quad \;\;\;{\mkern 1mu} 0 \le i < {H_{{\rm{out}}}},\;0 \le j < {W_{{\rm{out}}}},
\end{equation}
where \(X\) and \(Y\) represent the input data and the output of the max pooling operation, respectively, \(({k_h},~{k_w})\) is the kernel size, \(({s_h},~{s_w})\) is the stride, \(i\) and \(j\) denote the spatial indices of the output, \(a\) and \(b\) represent the max-pooling window offsets, and \({H_{{\rm{out}}}}\) and \({W_{{\rm{out}}}}\) denote the output dimensions, in which they can be determined by \({H_{{\rm{out}}}} = \left[ {\left( {H - {k_h}} \right)/{s_h}} \right] + 1,\)~\({W_{{\rm{out}}}} = \left[ {\left( {W - {k_w}} \right)/{s_w}} \right] + 1,\) where \(H\) and \(W\) denote the number of rows and columns of the input. Because a kernel size larger than 2 may overlook important local signal features, and a stride greater than 1 reduces the dimensions of the input data—potentially leading to a loss of useful information—we choose a kernel size of (2, 2) and a stride of (1, 1).

To mitigate cycle skipping caused by amplitude discrepancies between observed and simulated data, both datasets are typically first energy-normalized before being used to construct the loss function. Therefore, the new Euclidean loss can be written as:

\begin{equation}\label{eq7}
{\hat L_{Eucl}} = \sqrt {\sum\limits_{s,r,t} {{{\left( {{{\hat d}_{sim}}(s,r,t) - {{\hat d}_{obs}}(s,r,t)} \right)}^2}} } ,
\end{equation}
where \({{{\hat d}_{sim}}}\) and \({{{\hat d}_{obs}}}\) denote the energy-normalized forms of \({{d_{{\rm{sim}}}}}\) and \({{d_{{\rm{obs}}}}}\), respectively, whose mathematical expressions are given by \({{\hat d}_{{\rm{sim}}}} = \frac{{{d_{{\rm{sim}}}}}}{{{{\left\| {{d_{{\rm{sim}}}}} \right\|}_2}}}\), and \({{\hat d}_{{\rm{obs}}}} = \frac{{{d_{{\rm{obs}}}}}}{{{{\left\| {{d_{{\rm{obs}}}}} \right\|}_2}}}\).

The new HTE loss function can be expressed as:
\begin{equation}\label{eq8}
{\hat L_{HTE}} = \sqrt {\sum\limits_{s,r,t} {{{\left( {\hat e_{sim}^p(s,r,t) - \hat e_{obs}^p(s,r,t)} \right)}^2}} },
\end{equation}
where \({\hat e_{sim}} = \sqrt {\hat d_{sim}^2 + \hat d_{Hsim}^2} ,\)~\({\hat e_{obs}} = \sqrt {\hat d_{obs}^2 + \hat d_{Hobs}^2}\). Here, \({{\hat d}_{Hsim}}\) and \({{\hat d}_{Hobs}}\) denote the Hilbert-transformed counterparts of \({{\hat d}_{{\rm{sim}}}}\) and \({{\hat d}_{{\rm{obs}}}}\), respectively. 

In addition, the new MPBAE loss function is defined as follows:
\begin{equation}\label{eq9}
{\hat L_{MPBAE}} = \sqrt {\sum\limits_{s,r,t} {{{\left( {{{\hat \zeta }_{sim}}(s,r,t) - {{\hat \zeta }_{obs}}(s,r,t)} \right)}^2}} } ,
\end{equation}
where \({{{\hat \zeta }_{sim}}}\) and \({{{\hat \zeta }_{obs}}}\) are the MPBAEs obtained by applying max pooling to \({{\hat d}_{{\rm{sim}}}}\) and \({{\hat d}_{{\rm{obs}}}}\), respectively.

\begin{figure}
\centering
\includegraphics[width=1\textwidth]{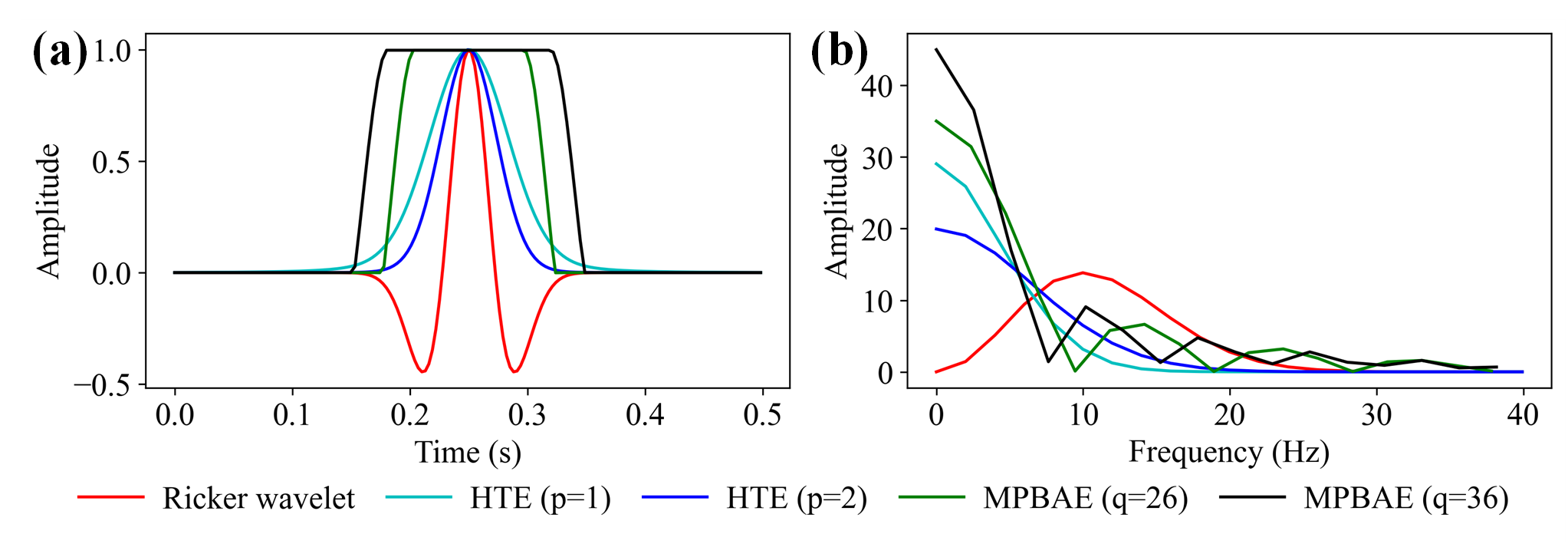}
\caption{Ricker wavelet along with its envelopes and spectrum. (a) The Ricker wavelet and envelopes obtained using different methods, (b) The spectra of the Ricker wavelet and its envelopes.}
\label{fig1}
\end{figure}

To demonstrate that MPBAE contains richer low-frequency components than HTE, we consider a Ricker wavelet with a dominant frequency of 10 Hz and compute its HTE and MPBAE, as shown in Fig. \ref{fig1}(a). The corresponding spectra are presented in Fig. \ref{fig1}(b). The wavelet has a time sampling interval of 3 ms and a total duration of 0.5 s. As demonstrated in Fig. \ref{fig1}, MPBAE exhibits stronger low-frequency content compared with HTE, and the magnitude of these low frequencies increases with the number of max-pooling levels \(q\). For HTE, using \(p\) = 2 instead of \(p\) = 1 reduces the strength of the low-frequency components. However, squaring the envelope enhances high-energy early-arrival components (e.g., direct waves) while suppressing low-energy late-arrival components (e.g., reflection waves), thereby reducing envelope complexity and helping FWI mitigate cycle skipping. In light of this, HTE-based inversion typically employs \(p\) = 2; this choice is also adopted in the numerical experiments presented in this study. 

\subsection{Shot patching and MPBAEP loss}
To derive the MPBAEP loss, we first consider the expression of the velocity gradient, which can be written based on the adjoint-state method \citep{plessix2006review} as:
\begin{equation}
\label{eq10}
\frac{{\partial L}}{{\partial {\bf{m}}}} = \int_0^T {\frac{2}{{{v^3}}}} \frac{{{\partial ^2}u}}{{\partial {t^2}}} \cdot {u^\bot }dt,
\end{equation}
where \({u^ \bot }\) is the adjoint wavefield, \(T\) denotes the recording time. Since the acoustic wave equation (\ref{eq1}) is self adjoint, the propagation of the adjoint wavefield is described by the following wave equation:

\begin{equation}
\label{eq11}
\frac{{{\partial ^2}{u^ \bot }({\bf{x}},t)}}{{\partial {t^2}}} = {v^2}({\bf{x}})\left( {\frac{{{\partial ^2}{u^ \bot }({\bf{x}},t)}}{{\partial {x^2}}} + \frac{{{\partial ^2}{u^ \bot }({\bf{x}},t)}}{{\partial {z^2}}}} \right) + \frac{{\partial L}}{{\partial {d_{sim}}}},
\end{equation}
where \(\partial L/\partial {{\bf{d}}^{{\rm{sim}}}}\) is the adjoint source. The adjoint source operator differs depending on the selected loss function. Specifically, for the Euclidean loss, the corresponding adjoint source operator can be expressed as: 
\begin{equation}
\label{eq12}
\frac{{\partial {L_{{\rm{Eucl}}}}}}{{\partial {d_{sim}}}} = \frac{{{d_{sim}} - {d_{obs}}}}{{{{\left\| {{d_{sim}} - {d_{obs}}} \right\|}_2}}}.
\end{equation}

From the adjoint source formula (\ref{eq12}), we observe that the adjoint source corresponding to the Euclidean loss involves energy normalization of the data residual. Motivated by this observation, we divide each shot gather into smaller patches and compute the data loss within each patch separately. For the Euclidean loss, this patch-based approach introduces an independent normalization for each patch during adjoint source computation. As a result, the adjoint source exhibits a more balanced energy distribution, leading to more balanced gradients. A non-overlapping patching strategy is adopted in this study, and it does not introduce discontinuities into the inversion results. This is because when the inverted velocity model approaches the true model, the difference between the observed and simulated data becomes nearly zero. Consequently, even after normalizing the data residual within each patch, the resulting residual remains close to zero. We adopt a two-dimensional patching strategy, with patch sizes defined along the spatial and temporal dimensions as s1 and s2, respectively. Since this process operates at the shot gather level, we refer to it as the shot patching strategy.

\begin{figure}
\centering
\includegraphics[width=1\textwidth]{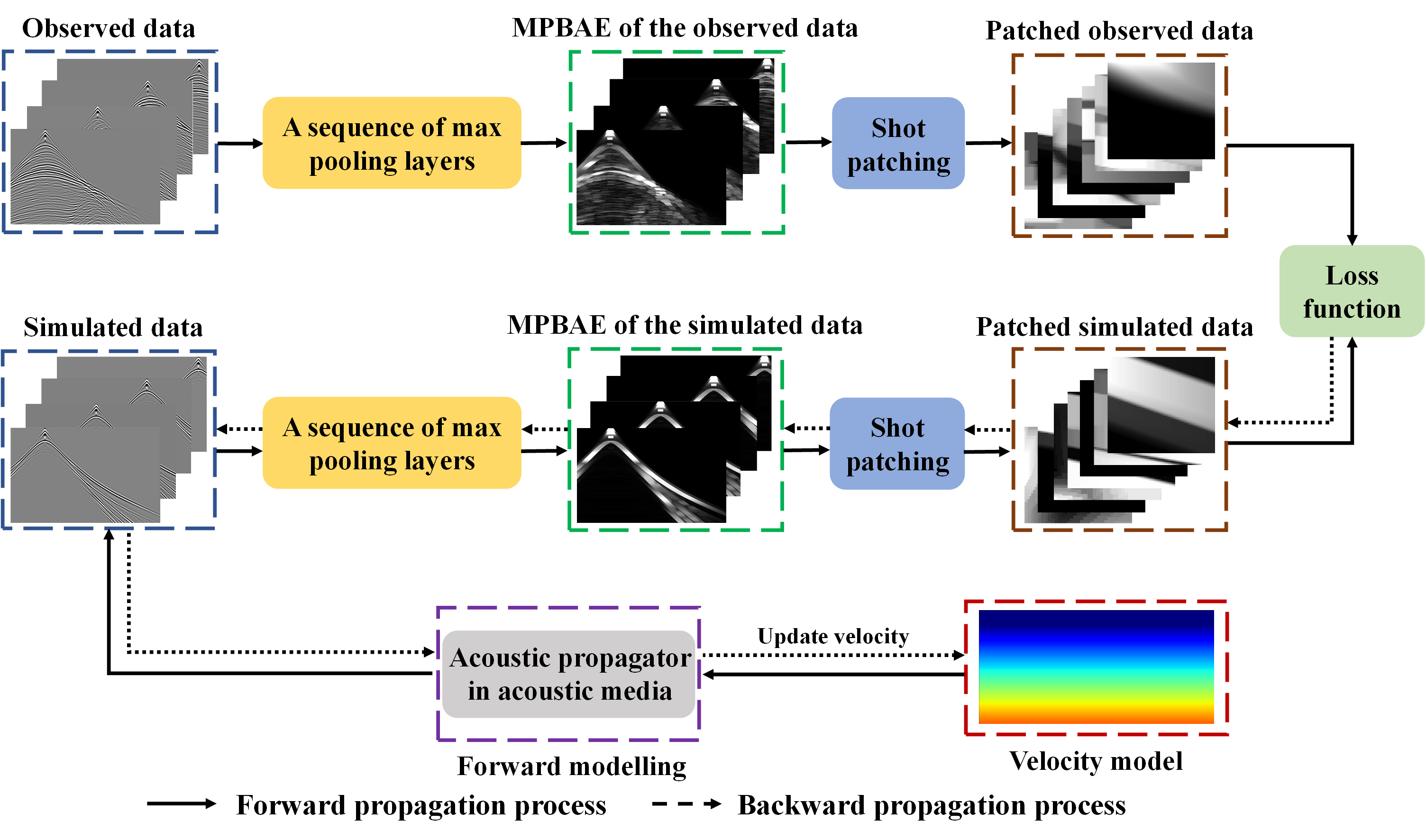}
\caption{Schematic of the proposed MPBAEP-FWI.}
\label{fig2}
\end{figure}

Fig. \ref{fig2} illustrates the schematic of the proposed MPBAEP-FWI, in which a series of max-pooling operations are applied to both the simulated and observed data to obtain approximate envelopes. The resulting approximate envelopes are then divided into multiple equal-sized patches, and the loss function is constructed based on these patches. In this study, the velocity gradient is computed using automatic differentiation, which yields results equivalent to those of the adjoint-state method \citep{richardson2018seismic}. 

\subsection{Convexity Analysis of the Misfit Function}
\begin{figure}
\centering
\includegraphics[width=0.8\textwidth]{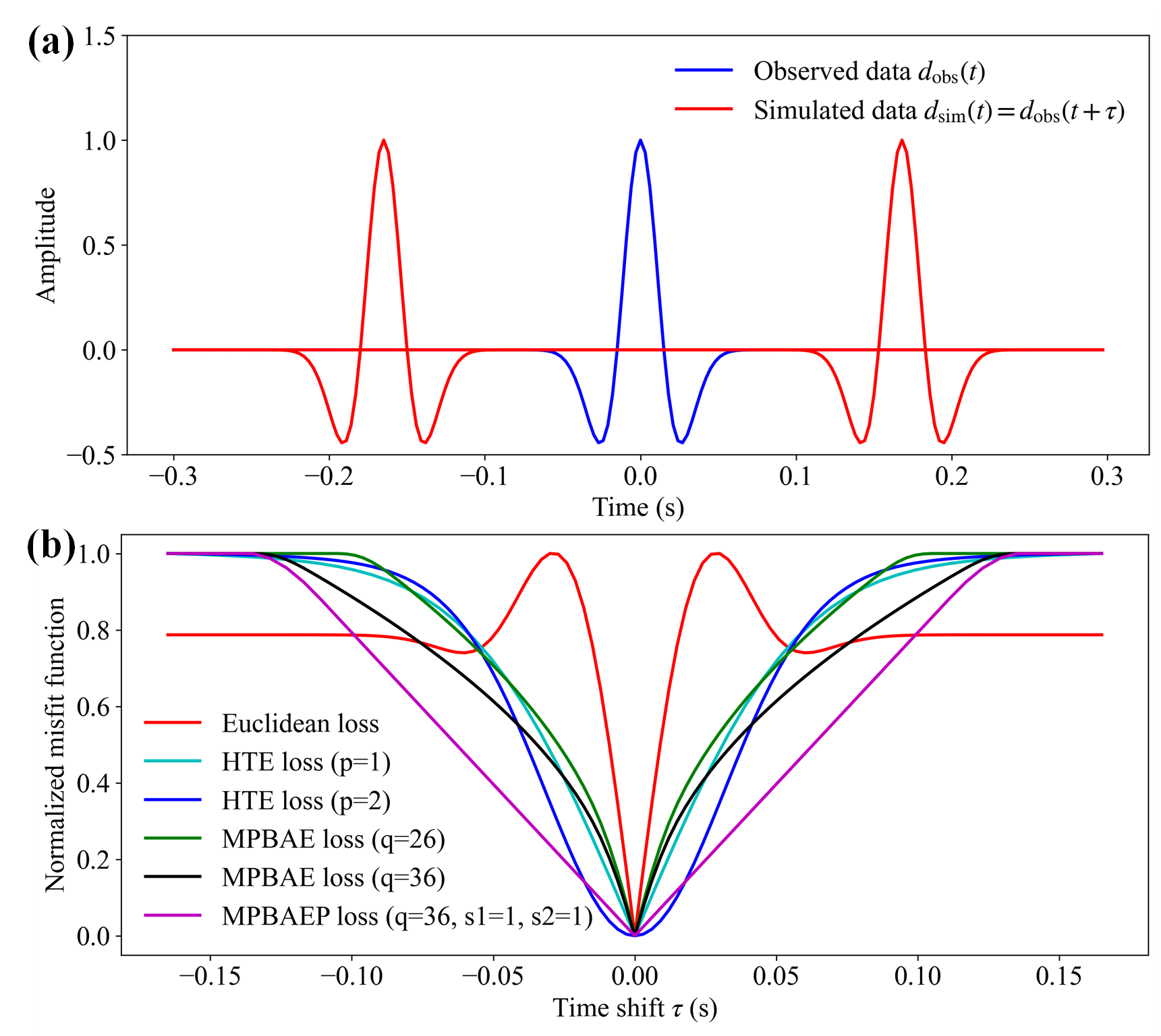}
\caption{Convexity analysis of the misfit functions (equations (\ref{eq2}), (\ref{eq4}), and (\ref{eq5})): (a) observed signal (blue) and simulated signals (red) with different time shifts, where \(\tau\) denotes the time shift varying from -0.165 to 0.165 s; (b) corresponding misfit function curves.}
\label{fig3}
\end{figure}

We first analyze the convexity properties of the Euclidean (equation (\ref{eq2})), HTE (equation (\ref{eq4})), MPBAE (equation (\ref{eq5})), and MPBAEP (equation (\ref{eq5}) combines the shot patching) losses. As shown in Fig. \ref{fig3}(a), the observed data \({d_{{\rm{obs}}}}(t){\rm{ }}\) is taken as a Ricker wavelet with a dominat frequency of 15 Hz, while the simulated data \({d_{{\rm{sim}}}}(t){\rm{ }}\) is a time-shifted version of the observed data. The comparisons of the different misfit functions are shown in Fig. \ref{fig3}(b). We can observe that the Euclidean loss exhibits local minima, while the HTE, MPBAE, and MPBAEP losses demonstrate improved convexity. For the HTE loss, when \(p\) = 2, its convexity slightly decreases compared to \(p\) = 1 due to the reduced strength of the low-frequency components. However, as discussed in \cite{wu2014seismic}, setting \(p\) = 2 emphasizes the envelope of the first arrivals while attenuating the envelope of reflected waves, which is more effective for mitigating cycle skipping. For the MPBAE loss, increasing \(q\) improves the convexity of the objective function and thus helps mitigate cycle skipping. However, overly large values of \(q\) may suppress useful signal information, resulting in degraded inversion accuracy. Compared with the HTE loss, the MPBAE loss exhibits better convexity when an appropriate \(q\) is chosen, providing a more effective mechanism to mitigate cycle skipping. It also admits a steeper drop near the global minimum, which helps achieve high resolution results. Furthermore, when the MPBAEP loss is adopted, we observe an improved convexity of the objective function compared to the MPBAE loss, which further helps mitigate cycle skipping. 

\begin{figure}
\centering
\includegraphics[width=0.8\textwidth]{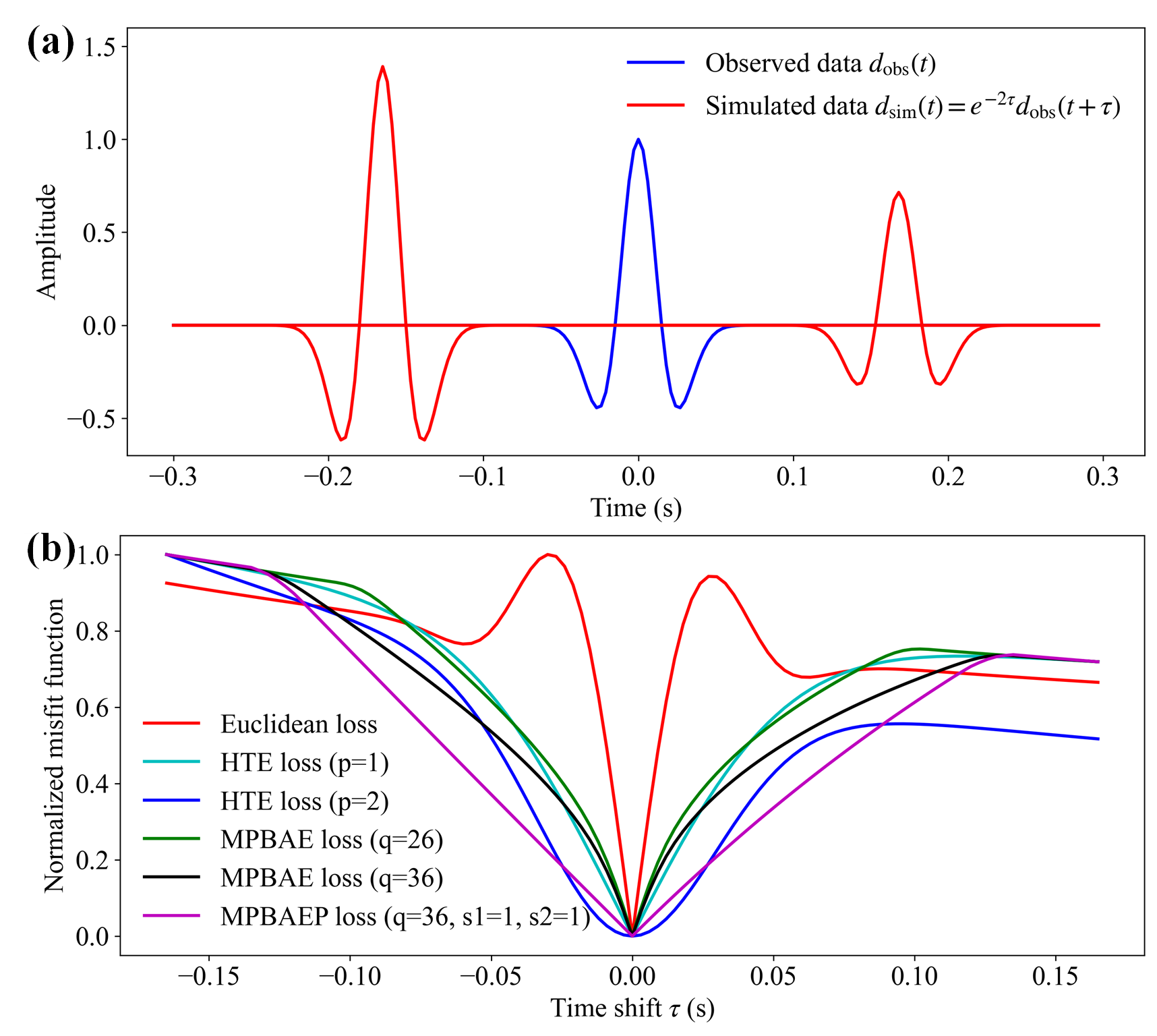}
\caption{Convexity analysis of the misfit functions(equations (\ref{eq2}), (\ref{eq4}), and (\ref{eq5})): (a) observed signal (blue) and simulated signals (red) with time shifts and amplitude variations, (b) the resulting misfit function curves.}
\label{fig4}
\end{figure}

\begin{figure}
\centering
\includegraphics[width=0.8\textwidth]{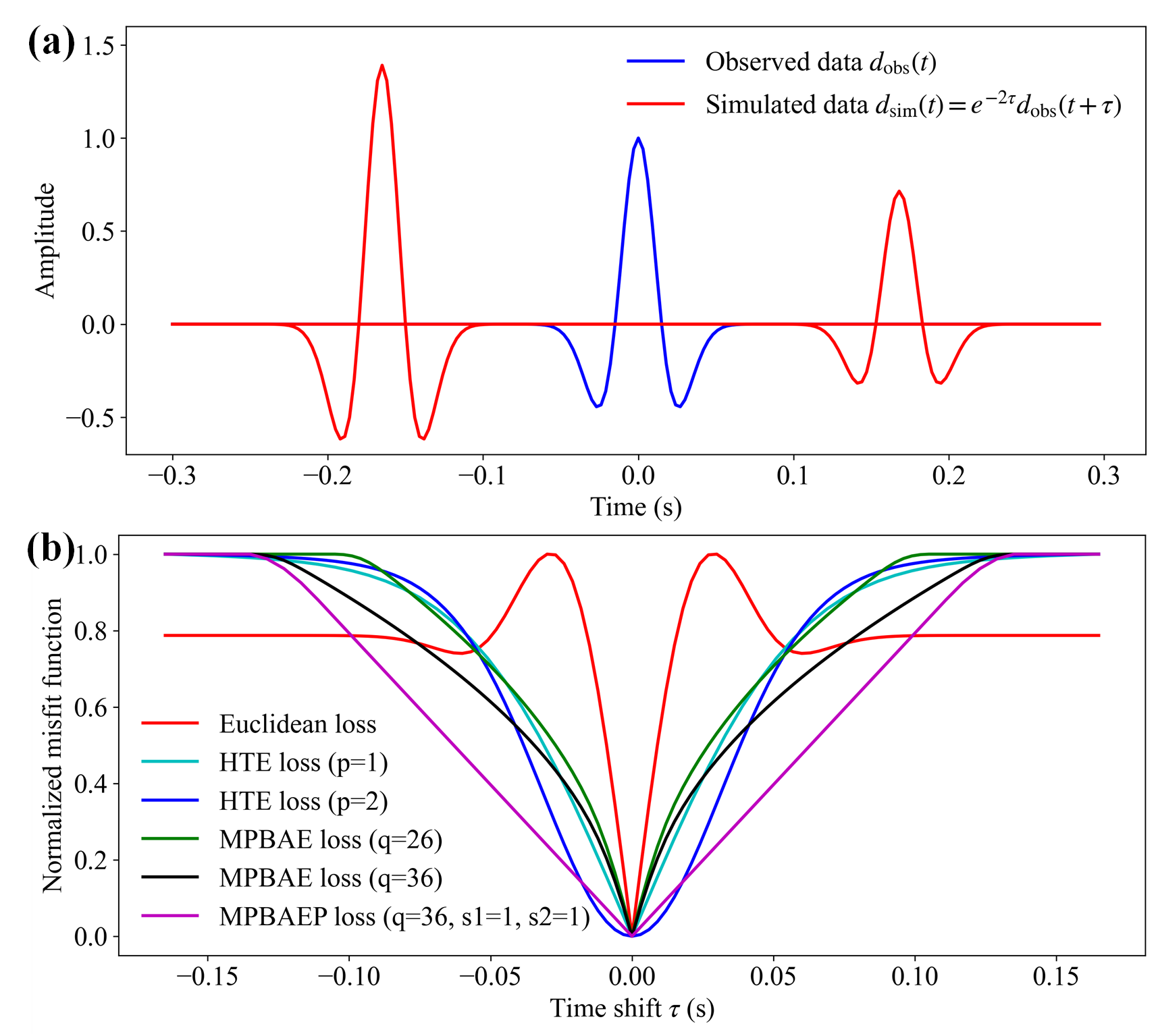}
\caption{Convexity analysis of the misfit functions(equations (\ref{eq7}), (\ref{eq8}), and (\ref{eq9})): (a) observed signal (blue) and simulated signals (red) with varying time shifts and amplitude changes; (b) the corresponding misfit function curves.}
\label{fig5}
\end{figure}

In practical FWI applications, the amplitude ranges of the observed and simulated data are not always consistent, which may induce cycle skipping due to amplitude discrepancies between them. As shown in Fig. \ref{fig4}, when both the traveltime and amplitude of the simulated data are varied simultaneously, the HTE (equation (\ref{eq4})), MPBAE (equation (\ref{eq5})), and MPBAEP (equation (\ref{eq5}) combines the shot patching) loss functions no longer exhibit convexity. When the observed and simulated data are each normalized by their respective energies, the loss function nonconvexity caused by amplitude discrepancies is corrected, thereby restoring convexity, as shown in Fig. \ref{fig5}.

\section{NUMERICAL EXPERIMENTS}
In this section, we use the Overthrust model, the Marmousi model, and a field dataset to demonstrate that MPBAE-FWI can mitigate cycle skipping. In addition, compared with MPBAE-FWI, MPBAEP-FWI further balances gradient energy and yields more accurate inversion results. For the synthetic data tests, a free-surface boundary condition is adopted. Furthermore, frequencies below 3 Hz are removed.

\subsection{Overthrust model}
\begin{figure*}
\centering
\includegraphics[width=1\textwidth]{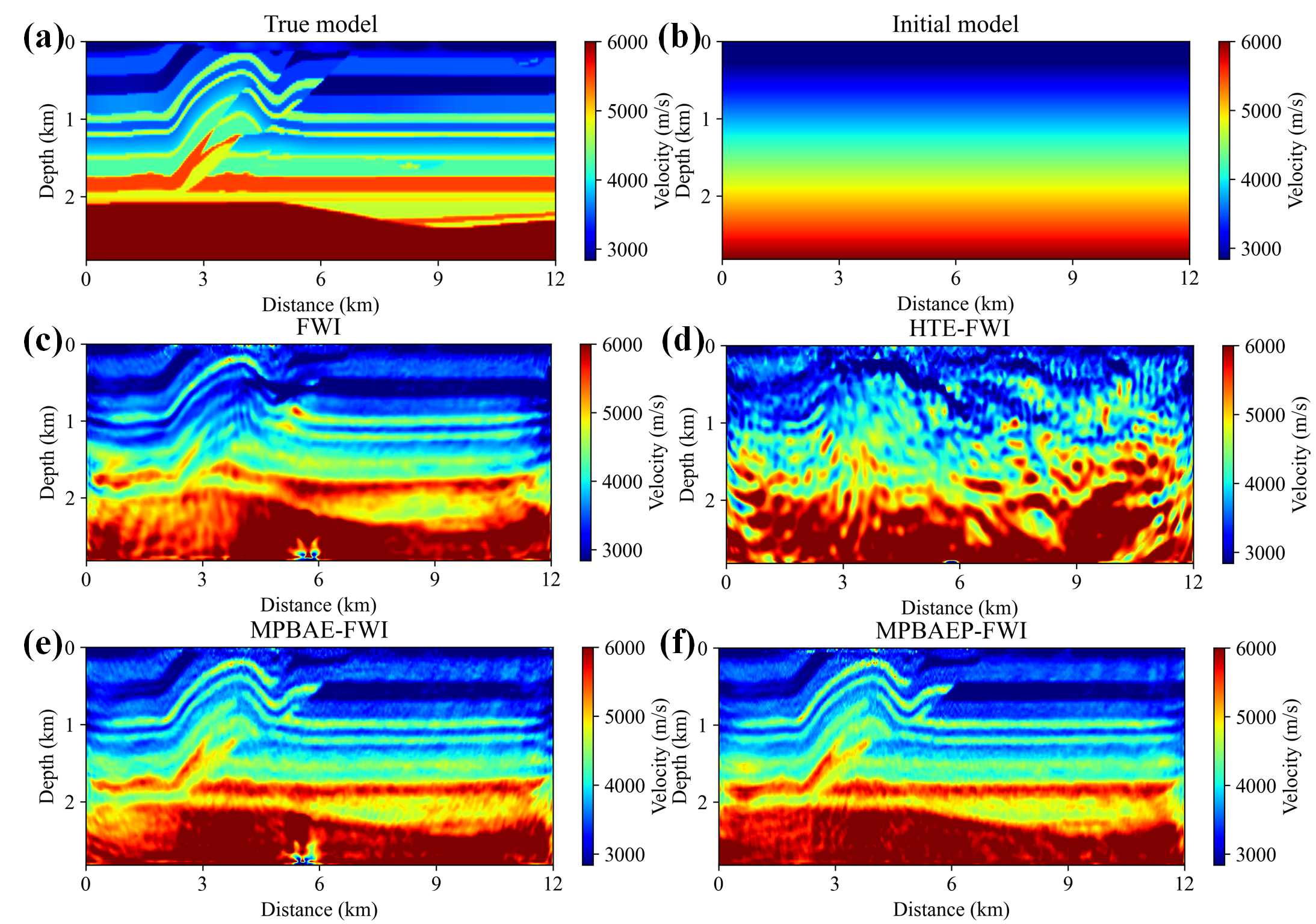}
\caption{Comparison of inversion results for the Overthrust model: (a) true Overthrust model, (b) initial velocity model, and velocity models obtained after 50 iterations using (c) FWI, (d) HTE-FWI, (e) MPBAE-FWI, and (f) MPBAEP-FWI.}
\label{fig6}
\end{figure*}

We first use the Overthrust model \citep{aminzadeh1994seg} to demonstrate the effectiveness of the proposed MPBAEP-FWI. Fig. \ref{fig6}(a) shows the true Overthrust model, which is represented on a 400 × 94 grid with a constant spacing of 30 m in both directions. The source wavelet is a Ricker wavelet with a peak frequency of 5 Hz. A total of 30 shots are evenly distributed along the surface at 390 m intervals, with the first source placed at the left boundary of the model. For each shot, seismic data are recorded by 400 receivers positioned at the surface, spaced uniformly at 30 m. Each shot is recorded for 6 seconds with a time step of 3 ms. Fig. \ref{fig6}(b) shows the initial velocity model used for FWI, which is a linearly increasing model with velocities ranging from 2500 m/s to 6000 m/s. 

Figs \ref{fig6}(c)–\ref{fig6}(f) show the inversion results after 50 iterations using the Euclidean loss (equation (\ref{eq7})), HTE loss (equation (\ref{eq8})), MPBAE loss (equation (\ref{eq9})), and MPBAEP loss (i.e., equation (\ref{eq9}) combined with shot patching), respectively. In this experiment, we use \(p\) = 2 for HTE-FWI, \(q\) = 10 for MPBAE-FWI, and \(q\) = 10 with patch sizes of s1 = s2 = 64 for MPBAEP-FWI. The Adam optimizer is employed with a step length of 40. As shown in Figs \ref{fig6}(c) and \ref{fig6}(d), the inversion results obtained using the Euclidean loss and HTE loss exhibit significant errors compared with the true velocity model. When the MPBAE loss is employed, an accurate inversion result is achieved, as we can see in Fig. \ref{fig6}(e). Furthermore, the use of the MPBAEP loss yields a more accurate velocity model, with particular improvement in the deep regions, as illustrated in Fig. \ref{fig6}(f).

\begin{figure*}
\centering
\includegraphics[width=1\textwidth]{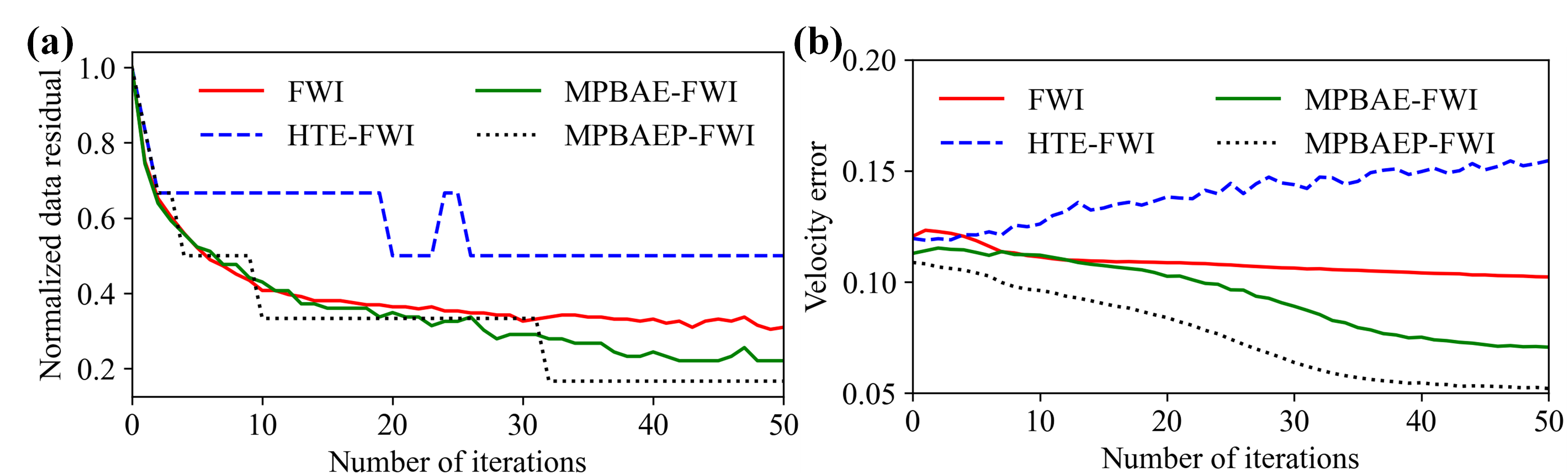}
\caption{(a) Convergence of normalized data residuals for different loss functions over iterations. (b) Evolution of velocity error over iterations for different loss functions.}
\label{fig7}
\end{figure*}

Fig. \ref{fig7}(a) shows the normalized data residual convergence curves for different loss functions in the Overthrust model test. We can observe that MPBAE-FWI outperforms both FWI and HTE-FWI in terms of convergence speed and final data residual. Moreover, MPBAEP-FWI further improves convergence efficiency over MPBAE-FWI and achieves the minimum final data misfit. Fig. \ref{fig7}(b) shows the variation of velocity error over iterations. It can be seen that MPBAE-FWI exhibits faster convergence than both FWI and HTE-FWI, and that MPBAEP-FWI further accelerates the convergence compared with MPBAE-FWI.

\begin{figure}
\centering
\includegraphics[width=1\textwidth]{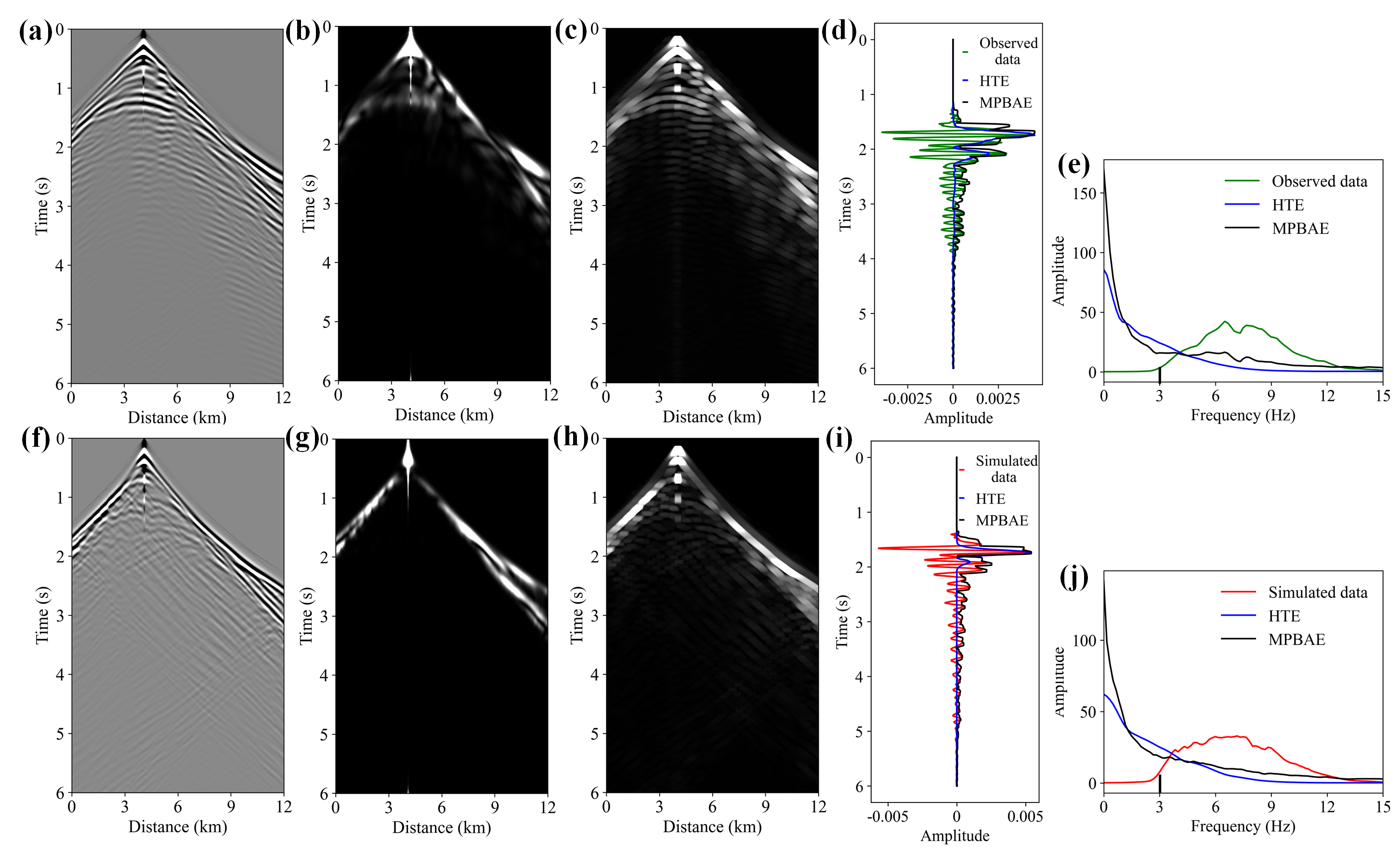}
\caption{Seismic data along with their envelopes and spectra. The first and second rows show the observed data and the simulated data from the initial model, respectively. Column 1: original data; Column 2: HTE; Column 3: MPBAE; Column 4: 1D profiles extracted from the first to the third columns at a horizontal distance of 8.1 km; Column 5: spectra of columns 1–3 (averaged over all traces).}
\label{fig8}
\end{figure}

The original data, together with the computed HTE and MPBAE and their corresponding spectra, are shown in Fig. \ref{fig8}. We can observe that MPBAE contains stronger low-frequency components than HTE, which helps more effectively mitigate cycle skipping. In addition, MPBAE better preserves the envelopes of both direct and reflected waves, thereby enabling higher-resolution inversion results.

\begin{figure}
\centering
\includegraphics[width=1\textwidth]{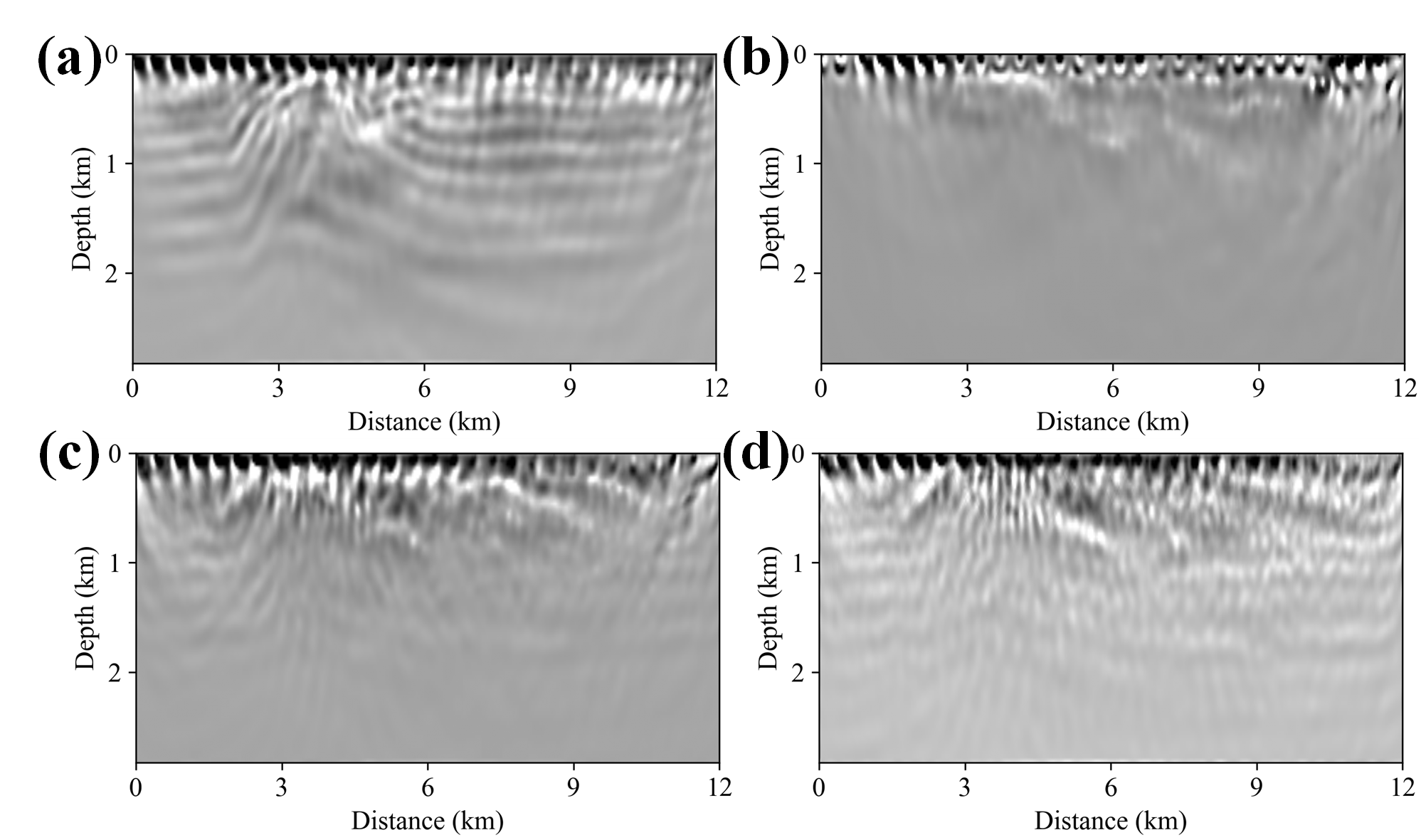}
\caption{Gradients of the Overthrust model after the first iteration using (a) Euclidean loss, (b) HTE loss, (c) MPBAE loss, and (d) MPBAEP loss.}
\label{fig9}
\end{figure}

Fig. \ref{fig9} shows the gradients after the first iteration obtained using the Euclidean loss, HTE loss, MPBAE loss, and MPBAEP loss. The reflectivity structures in the gradients obtained with the Euclidean loss and HTE loss exhibit noticeable depth deviations from the true subsurface interfaces, which helps explain their inability to recover accurate velocity models. In contrast, the gradient obtained with the MPBAE loss more accurately delineates the subsurface layering, indicating that the proposed method can effectively mitigate cycle skipping and provide a reliable search direction for velocity inversion. Furthermore, we can observe that the gradient computed with the MPBAEP loss exhibits a more balanced energy distribution compared with that obtained using the MPBAE loss, which helps explain its improved inversion accuracy.

Table \ref{tab1} summarizes the signal-to-noise ratio (SNR), the structural similarity index measure (SSIM), the root mean square error (RMSE), and the running time values for velocity models obtained with different loss functions, compared to the true velocity model. The inversion results obtained using the MPBAEP loss achieve the highest SNR and SSIM values and the lowest RMSE, while incurring only a marginal increase in computational cost compared with the other loss functions. This demonstrates that, compared with alternative objective functions, the MPBAEP loss can effectively mitigate cycle skipping and improve inversion accuracy with only a negligible additional computational burden.

\begin{table}
\centering 
\caption{Comparison of FWI accuracy using different loss functions for the Overthrust model. The computations are performed using a single NVIDIA A100 GPU. For each metric, the highest score is indicated in bold.}
\begin{tabular}{@{}cccccc@{}}
\hline
Method  & SNR (dB) & SSIM   & RMSE (km/s) & FWI iterations & Running time (s) \\ \hline
FWI     & 19.7966  & 0.4322 & 0.1023    & 50            & 190             \\
HTE-FWI  & 16.2880  & 0.1449 & 0.1533    & 50            & 191             \\
MPBAE-FWI  & 22.9794  & 0.5785 & 0.0709    & 50            & 209             \\
MPBAEP-FWI & \textbf{25.9080}  & \textbf{0.6461} & \textbf{0.0507}   & 50            & 213             \\ 
\hline
\end{tabular}
\label{tab1}
\end{table}

\subsection{Marmousi model}
\begin{figure*}
\centering
\includegraphics[width=\textwidth]{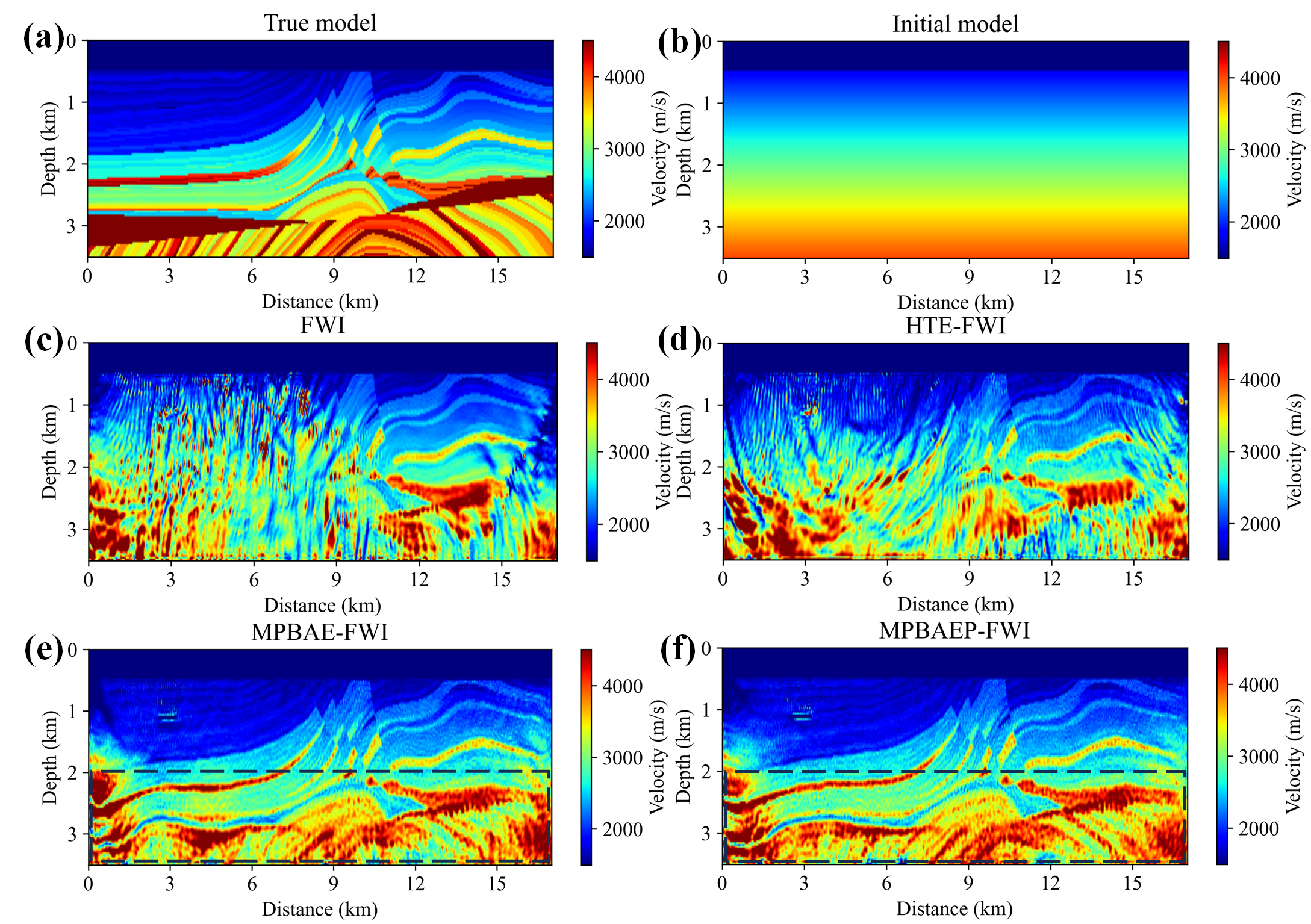}
\caption{Comparison of inversion results for the Marmousi model: (a) true Marmousi model, (b) initial velocity model, and inverted models after 200 iterations using (c) FWI, (d) HTE-FWI, (e) MPBAE-FWI, and (f) MPBAEP-FWI.}
\label{fig10}
\end{figure*}

We then use the complex Marmousi model \citep{martin2006marmousi2} to demonstrate the effectiveness of the proposed MPBAEP-FWI. As shown in Fig. \ref{fig10}(a), the Marmousi model is discretized into 567 × 117 grid points, with each grid cell measuring 30×30 \({{\rm{m}}^{\rm{2}}}\). The acquisition geometry includes 30 shot positions evenly spaced along the surface at an interval of 580 m. For every shot, 567 receivers are deployed uniformly along the surface, at 30 m spacing. The seismic source is a Ricker wavelet with a dominant frequency of 5 Hz. Each shot gather spans 6 s in duration and is sampled in time every 3 ms. Fig. \ref{fig10}(b) shows the initial velocity model, which is a linearly increasing model with velocity gradually increasing from the shallow to the deep layers.

The inversion results obtained using conventional FWI (equation (\ref{eq7})), HTE-FWI (equation (\ref{eq8})), MPBAE-FWI (equation (\ref{eq9})), and MPBAEP-FWI (i.e., equation (\ref{eq9}) combined with shot patching) are shown in Figs \ref{fig10}(c)–\ref{fig10}(f), respectively. In this test, the parameters are set to \(p\) = 2 for HTE-FWI, \(q\) = 18 for MPBAE-FWI, and \(q\) = 18 with patch sizes s1 = s2 = 64 for MPBAEP-FWI. The inversion is carried out using the Adam optimizer with a step length of 40. As we can see in Fig. \ref{fig10}, conventional FWI and HTE-FWI both deviate significantly from the true model. By contrast, MPBAE-FWI produces a result closely matching the true model. Moreover, MPBAEP-FWI further enhances inversion accuracy relative to MPBAE-FWI, especially in the deeper region marked by the black rectangular box.

\begin{figure*}
\centering
\includegraphics[width=\textwidth]{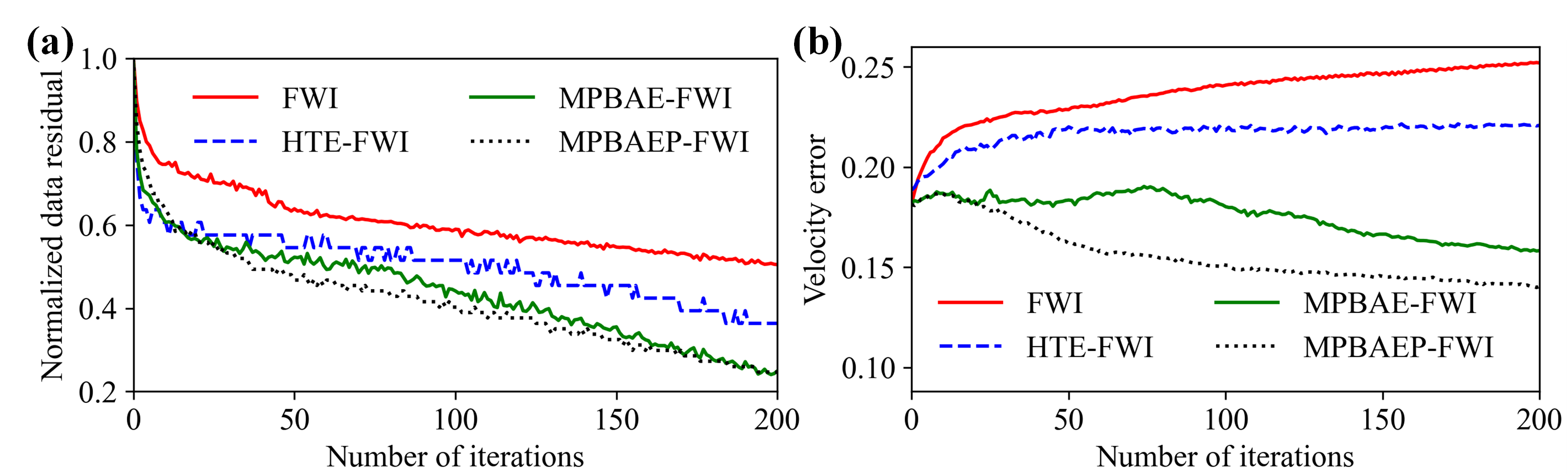}
\caption{Comparison of convergence performance for different FWI methods on the Marmousi model: (a) data residual convergence curves, and (b) velocity error convergence curves.}
\label{fig11}
\end{figure*}

Figs \ref{fig11}(a) and \ref{fig11}(b) show the normalized data residual convergence curves and velocity error convergence curves for the Marmousi model. Compared with the Euclidean loss and HTE loss, the MPBAE loss achieves faster convergence in terms of both data residuals and velocity errors, while also reaching lower final error values. Furthermore, compared with the MPBAE loss, the MPBAEP loss further accelerates convergence and yields even smaller final errors.

\begin{figure}
\centering
\includegraphics[width=1\textwidth]{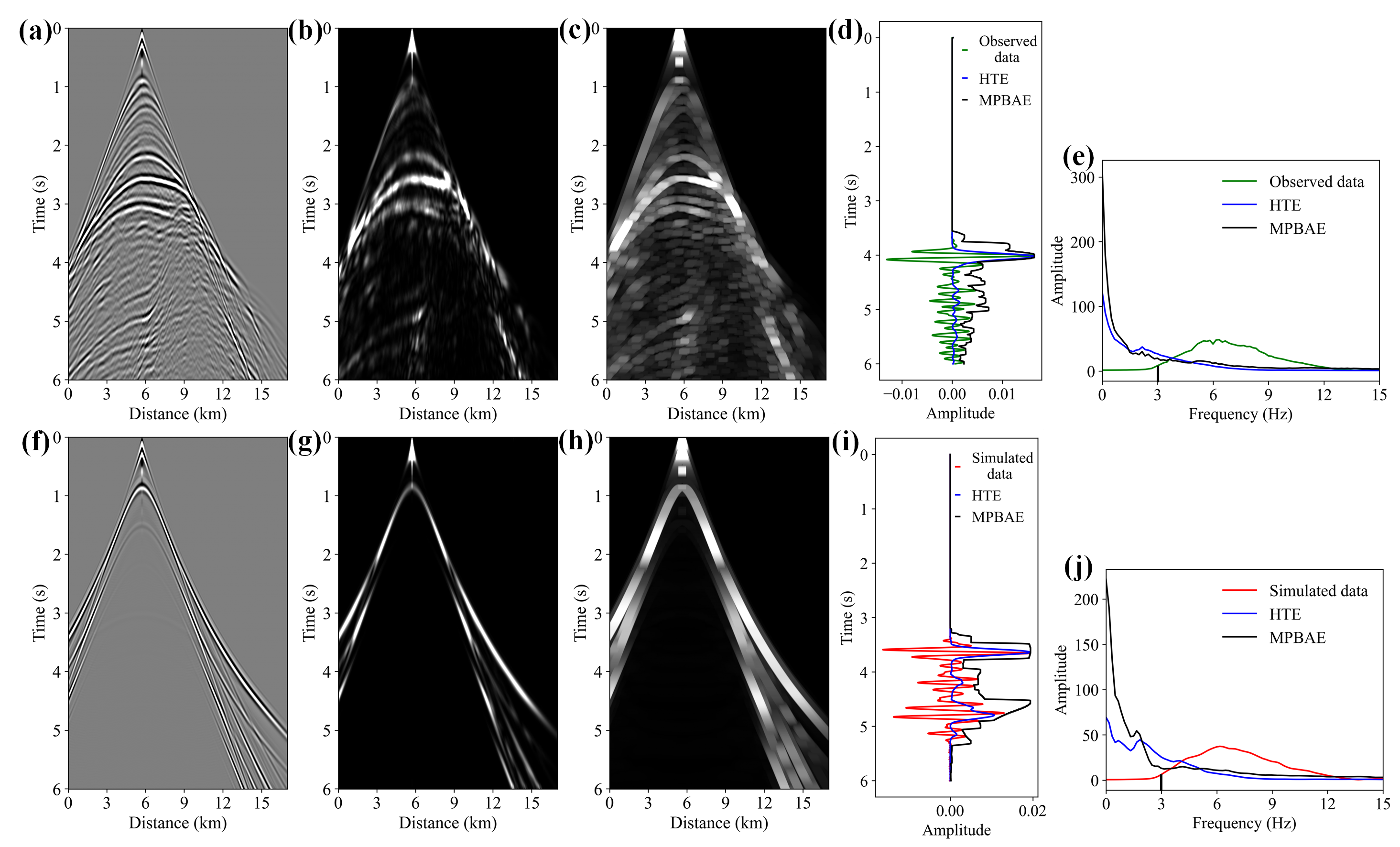}
\caption{Seismic data together with their envelopes and spectra. The first and second rows show the observed data and the simulated data generated from the initial model, respectively. Column 1: original data; Column 2: HTE; Column 3: MPBAE; Column 4: 1D profiles extracted from columns 1–3 at a horizontal distance of 12 km; Column 5: spectra of columns 1–3, calculated by averaging over all traces.}
\label{fig12}
\end{figure}

Fig. \ref{fig12} presents the original data together with the corresponding HTE, MPBAE, and their spectra. We can observe that MPBAE contains stronger low-frequency components than HTE, thereby providing greater capability to mitigate cycle skipping. Moreover, MPBAE better preserves the envelopes of both direct and reflected waves, which is beneficial for obtaining higher-resolution inversion results.

\begin{figure}
\centering
\includegraphics[width=1\textwidth]{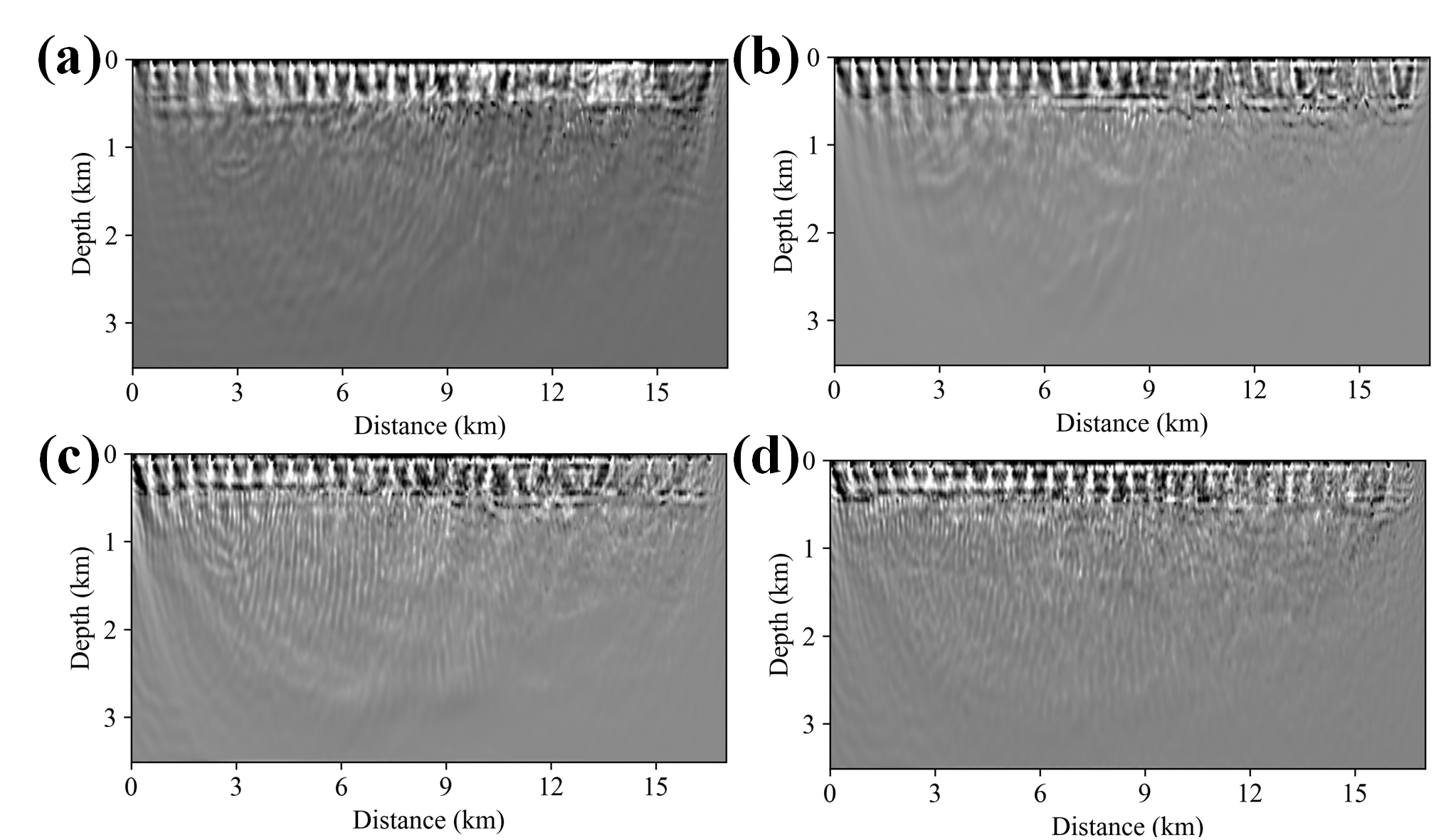}
\caption{Comparison of gradients for the Marmousi model after the first iteration using (a) Euclidean loss, (b) HTE loss, (c) MPBAE loss, and (d) MPBAEP loss.}
\label{fig13}
\end{figure}

The gradients after the first iteration obtained using the Euclidean loss, HTE loss, MPBAE loss, and MPBAEP loss are shown in Fig. \ref{fig13}. The seismic layers in the gradients produced by the Euclidean loss and HTE loss clearly deviate in depth from the true subsurface structures, which leads to their inability to recover an accurate velocity model. Owing to the complexity of the Marmousi model, the seismic layer features in the gradient obtained with the MPBAE loss are not easily interpretable, making it difficult to directly relate them to its inversion performance. Nevertheless, the gradient generated by the MPBAEP loss shows a more balanced amplitude distribution compared with that of the MPBAE loss, which accounts for its improved inversion accuracy.

Table \ref{tab2} summarizes the quantitative evaluation of the inverted velocity models using SNR, SSIM, and RMSE metrics with respect to the true velocity model. MPBAEP-FWI achieves the most accurate inversion results among all FWI methods, while incurring only a marginal increase in computational cost.

\begin{table}
\centering 
\caption{Assessment of FWI accuracy with various loss functions on the Marmousi model. All computations are carried out on a single NVIDIA A100 GPU. The best performance for each metric is highlighted in bold.}
\begin{tabular}{@{}cccccc@{}}
\hline
Method  & SNR (dB) & SSIM   & RMSE (km/s) & FWI iterations & Running time (s) \\ \hline
FWI     & 11.9681  & 0.3644 & 0.2521    & 200            & 853             \\
HTE-FWI  & 13.1329  & 0.3027 & 0.2204    & 200            & 974             \\
MPBAE-FWI  & 16.0234  & 0.5136 & 0.1581    & 200            & 1016             \\
MPBAEP-FWI & \textbf{17.0808}  & \textbf{0.5490} & \textbf{0.1399}   & 200            & 1033             \\ 
\hline
\end{tabular}
\label{tab2}
\end{table}

\subsection{Field data}
The field dataset used in this study was acquired offshore northwestern Australia and consists of 116 marine shots. An airgun array serves as the seismic source, and a 324-channel towed streamer was deployed for data acquisition. The hydrophones are evenly spaced at 25 m, and the shots were spaced approximately 90 m apart, covering a total lateral distance of about 20 km. Each shot record has a duration of 7 s and is originally sampled at 1 ms; the data are subsequently resampled to 2 ms for processing. Only frequency components below 10 Hz are used in the inversion process. In addition, frequencies below 2 Hz are missing, and the 2–4 Hz components are very weak. The grid spacing of the model is 25 m. For each shot, the source wavelet is estimated from the near-offset direct arrivals, assuming a seawater velocity of 1500 m/s.

\begin{figure*}
\centering
\includegraphics[width=1\textwidth]{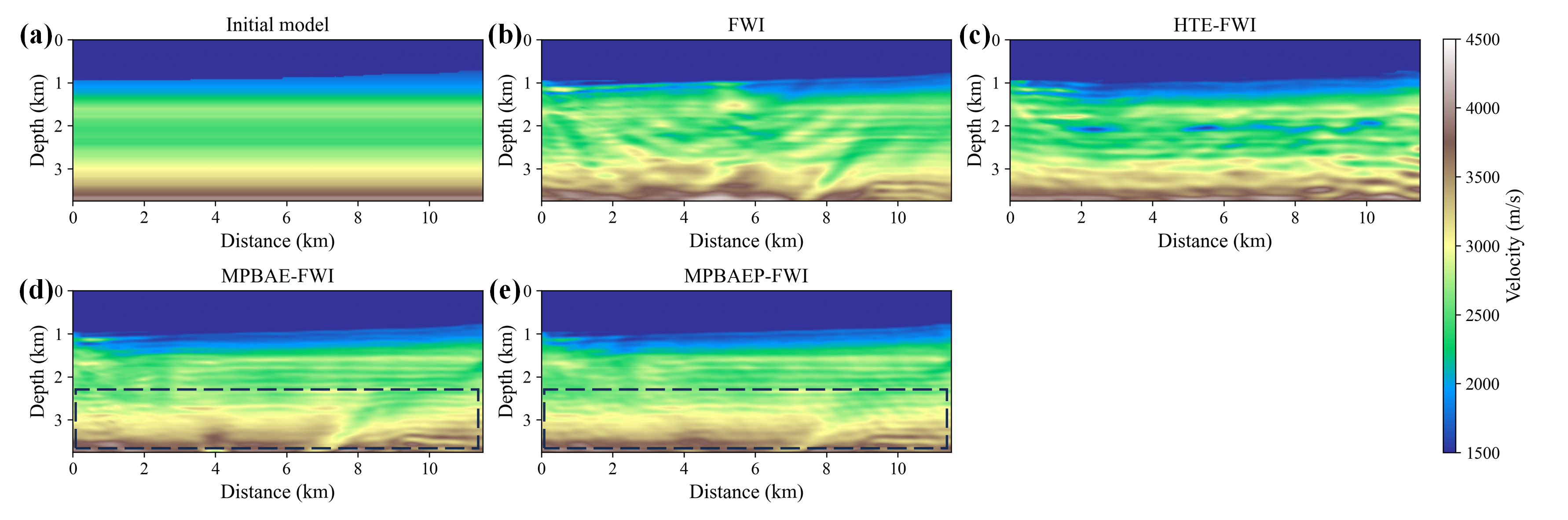}
\caption{Comparison of inversion results from different FWI methods applied to the Western Australia dataset: (a) initial velocity model; (b), (c), (d), and (e) show the results obtained from FWI, HTE-FWI, MPBAE-FWI, and MPBAEP-FWI, respectively.}
\label{fig14}
\end{figure*}

A horizontally layered velocity model is used as the initial model for the inversion, as shown in Fig. \ref{fig14}(a). We then apply different FWI methods to the data. The corresponding inversion results after 40 iterations are shown in Figs \ref{fig14}(b)-\ref{fig14}(e). In this test, the parameter \(p\) is set to 2 for HTE-FWI (equation (\ref{eq8})), \(q\) is set to 4 for MPBAE-FWI (equation (\ref{eq9})), and \(q\) = 4 with patch sizes s1 = s2 = 64 for MPBAEP-FWI (i.e., equation (\ref{eq9}) combined with shot patching). As we can see in Figs. \ref{fig14}(b) and \ref{fig14}(c), the results obtained using conventional FWI and HTE-FWI suffer from local minima, manifested as spurious low-velocity anomalies. In contrast, MPBAE-FWI yields a more accurate inversion result. Furthermore, MPBAEP-FWI achieves improved accuracy compared with MPBAE-FWI, particularly in the deeper regions of the model, as highlighted by the rectangular box.

\begin{figure}
\centering
\includegraphics[width=1\textwidth]{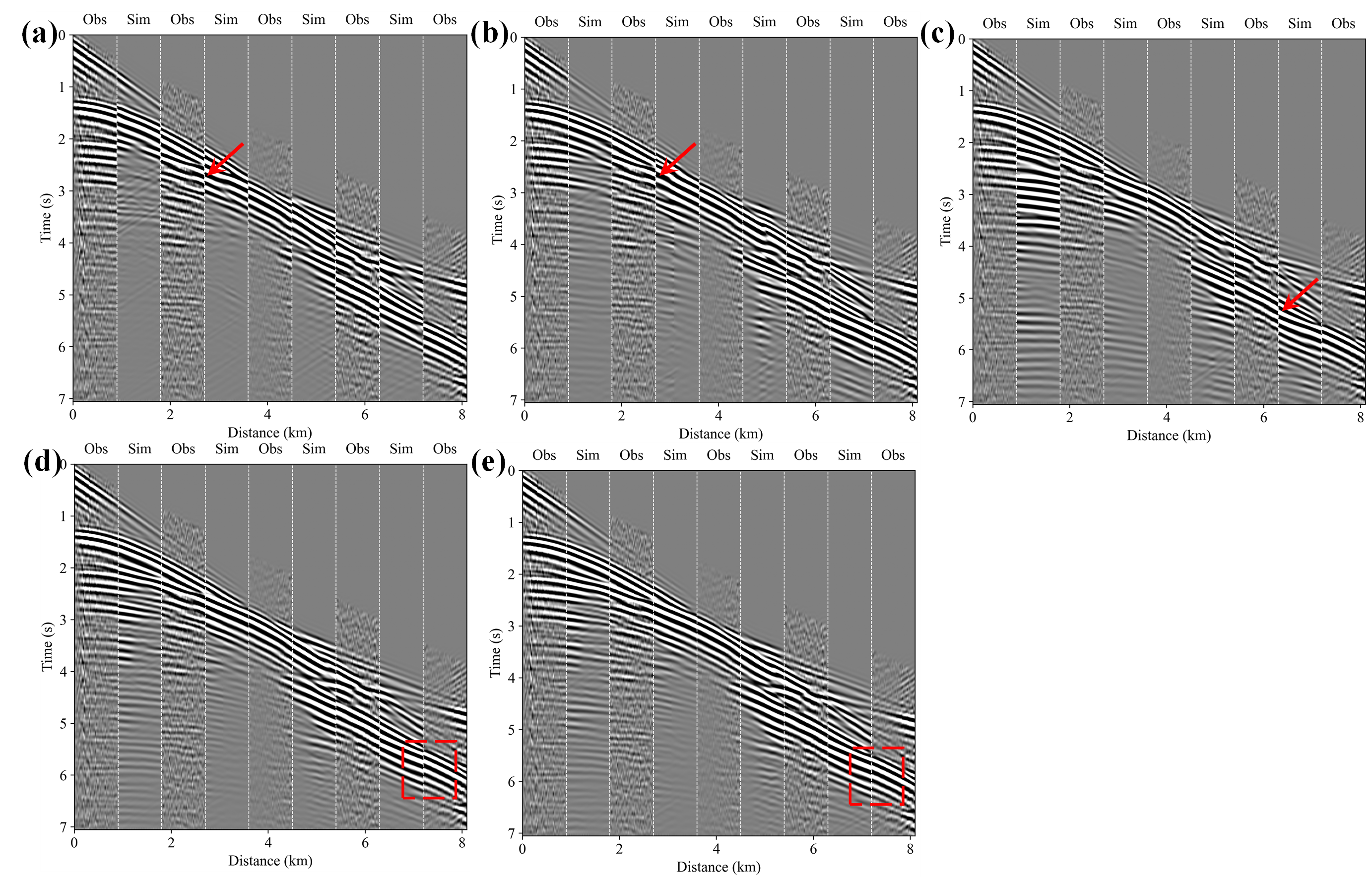}
\caption{Comparison between the observed and synthetic data generated by (a) the initial model and the inversion results obtained using (b) FWI, (c) HTE-FWI, (d) MPBAE-FWI, and (e) MPBAEP-FWI.}
\label{fig15}
\end{figure}

As shown in Fig. \ref{fig15}, the synthetic data generated from the inversion result of MPBAE-FWI match the observed data without noticeable cycle skipping. In contrast, the synthetic data generated from the initial model, FWI, and HTE-FWI inversion results exhibit clear cycle skipping relative to the observed data, as indicated by the red arrows. This demonstrates that MPBAE-FWI can mitigate cycle skipping and provides a more accurate inversion result. Furthermore, as highlighted by the red rectangular box, the synthetic data generated using MPBAEP-FWI show a better agreement with the observed data than those obtained using MPBAE-FWI, indicating that MPBAEP-FWI can further improve inversion accuracy compared with MPBAE-FWI.

\begin{figure*}
\centering
\includegraphics[width=1\textwidth]{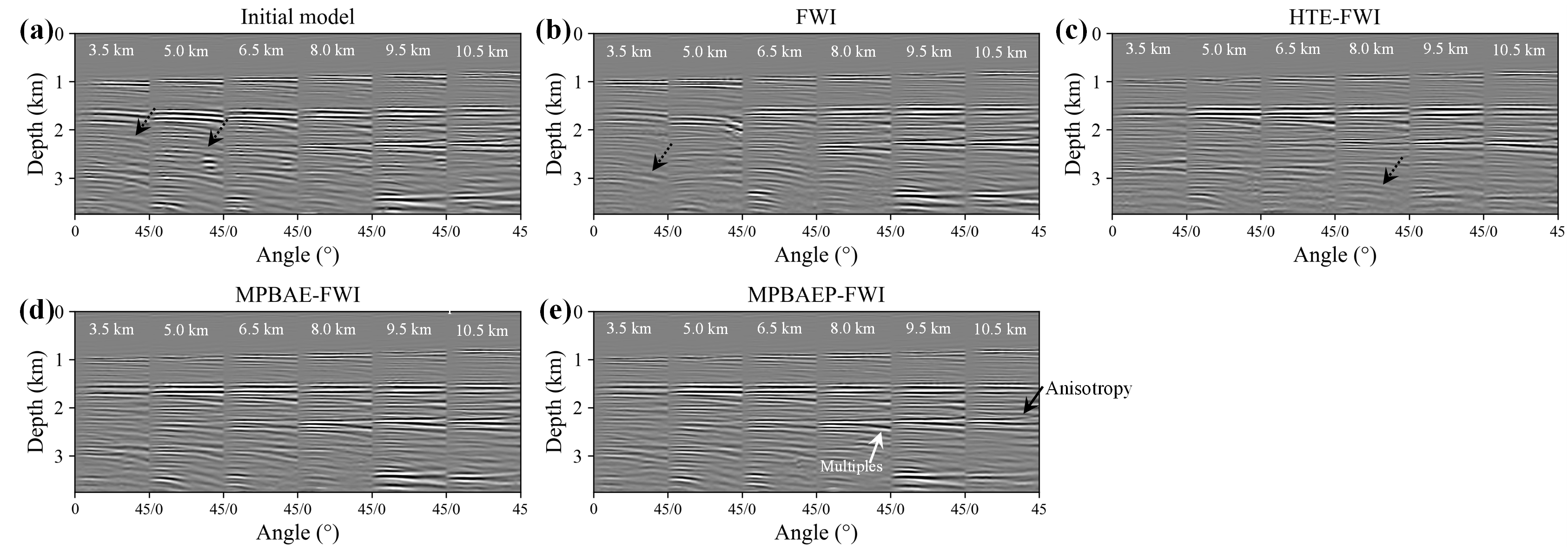}
\caption{Angle-domain common-image gathers at different locations computed using the initial model and the velocity models inverted by FWI, HTE-FWI, MPBAE-FWI, and MPBAEP-FWI. Each gather is displayed for reflection angles up to 45°.}
\label{fig17}
\end{figure*}

Fig. \ref{fig17} shows the angle-domain common-image gathers (ADCIGs) obtained using the velocity model presented in Fig. \ref{fig14}. The non-flat ADCIGs, indicated by the black dashed arrows in Figs \ref{fig17}(a)–\ref{fig17}(c), are caused by inaccuracies in the velocity model. This indicates that the velocity models obtained using FWI and HTE-FWI contain errors. As shown in Figs \ref{fig17}(d) and \ref{fig17}(e), the ADCIGs generated from the velocity models inverted by MPBAE-FWI and MPBAEP-FWI exhibit well-flattened events, demonstrating that both methods are capable of producing accurate velocity models. As shown in Fig. \ref{fig17}(e), the downward-curved gather indicated by the white solid arrow is associated with multiple reflection. The upward curvature observed at large angles, highlighted by the black solid arrows, may result from anisotropic effects \citep{mu2026full}. The inverted velocity model in Fig. \ref{fig14} indicates predominantly horizontal stratification in this region, suggesting the possible presence of vertical transverse isotropy (VTI). This provides a plausible explanation for the upward-curved events observed in the angle-domain image gathers.

\section{DISCUSSION}

\subsection{Effect of pooling depth (\(q\)) on FWI performance}
\begin{figure}
\centering
\includegraphics[width=1\textwidth]{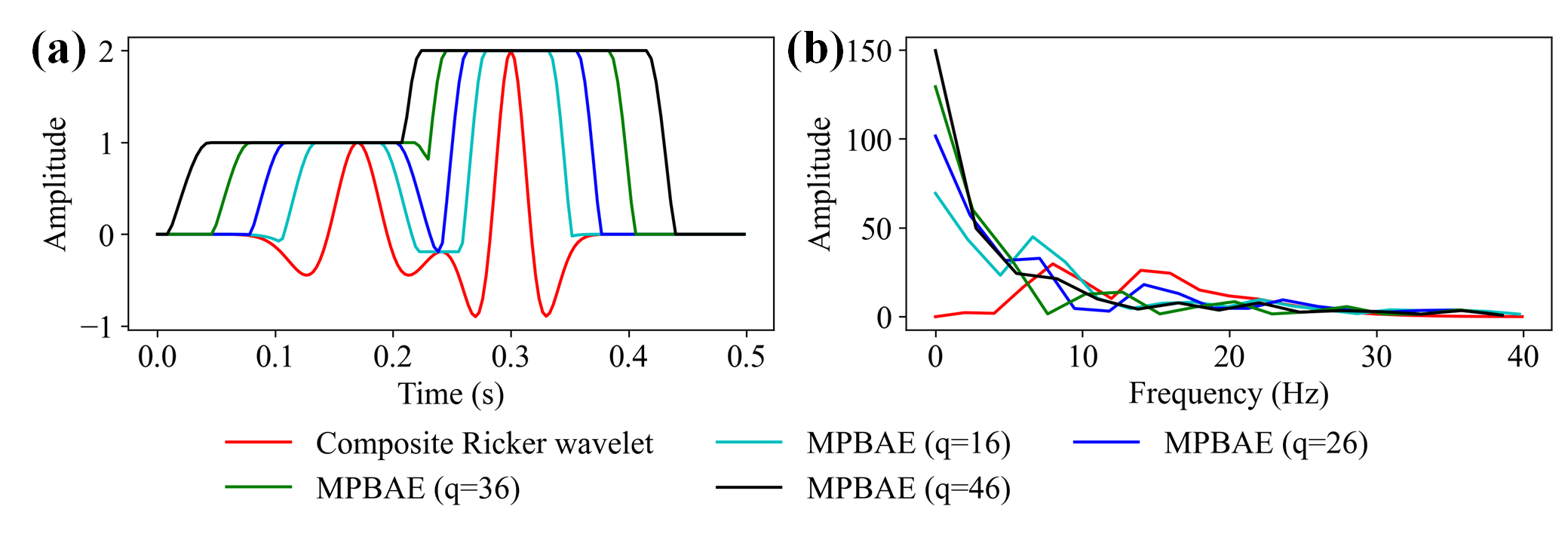}
\caption{Composite Ricker wavelet and its corresponding MPBAEs and spectra. (a) Composite Ricker wavelet together with MPBAEs for different pooling depths \(q\). (b) Frequency spectra of the composite Ricker wavelet and its MPBAEs.}
\label{figd1}
\end{figure}

In the theoretical analysis (Figs \ref{fig1} and \ref{fig3}), we use a Ricker wavelet to investigate the effect of increasing pooling depth \(q\). The results show that the resulting MPBAE contains progressively stronger low-frequency components, which is beneficial for mitigating cycle skipping. However, for complex signals, the low-frequency content of the MPBAE also increases as \(q\) increases. Beyond a certain critical point, further increasing \(q\) begins to distort the envelope structure, causing multiple wave packets (or events) to merge into a single one, which is unfavorable for inversion accuracy. To illustrate this more intuitively, in Fig. \ref{figd1}, we consider a composite Ricker wavelet signal and compute MPBAEs with different values of \(q\). When \(q{\rm{ <  26}}\), two distinct wave packets can be clearly observed, corresponding to the two individual Ricker wavelets. However, when \(q\) = 36, although the corresponding spectrum contains stronger low-frequency components, the MPBAE collapses into a single wave packet, which is inconsistent with the true two-event structure of the signal. Therefore, in practical inversion, we suggest extracting a representative seismic trace and applying max pooling with progressively increasing \(q\). The optimal pooling depth is achieved when the resulting envelope exhibits sufficiently strong low-frequency content while preserving the essential structure of the original signal. This \(q\) value is then selected as the optimal choice.

\subsection{Impact of patch size on inversion results}
\begin{figure*}
\centering
\includegraphics[width=1\textwidth]{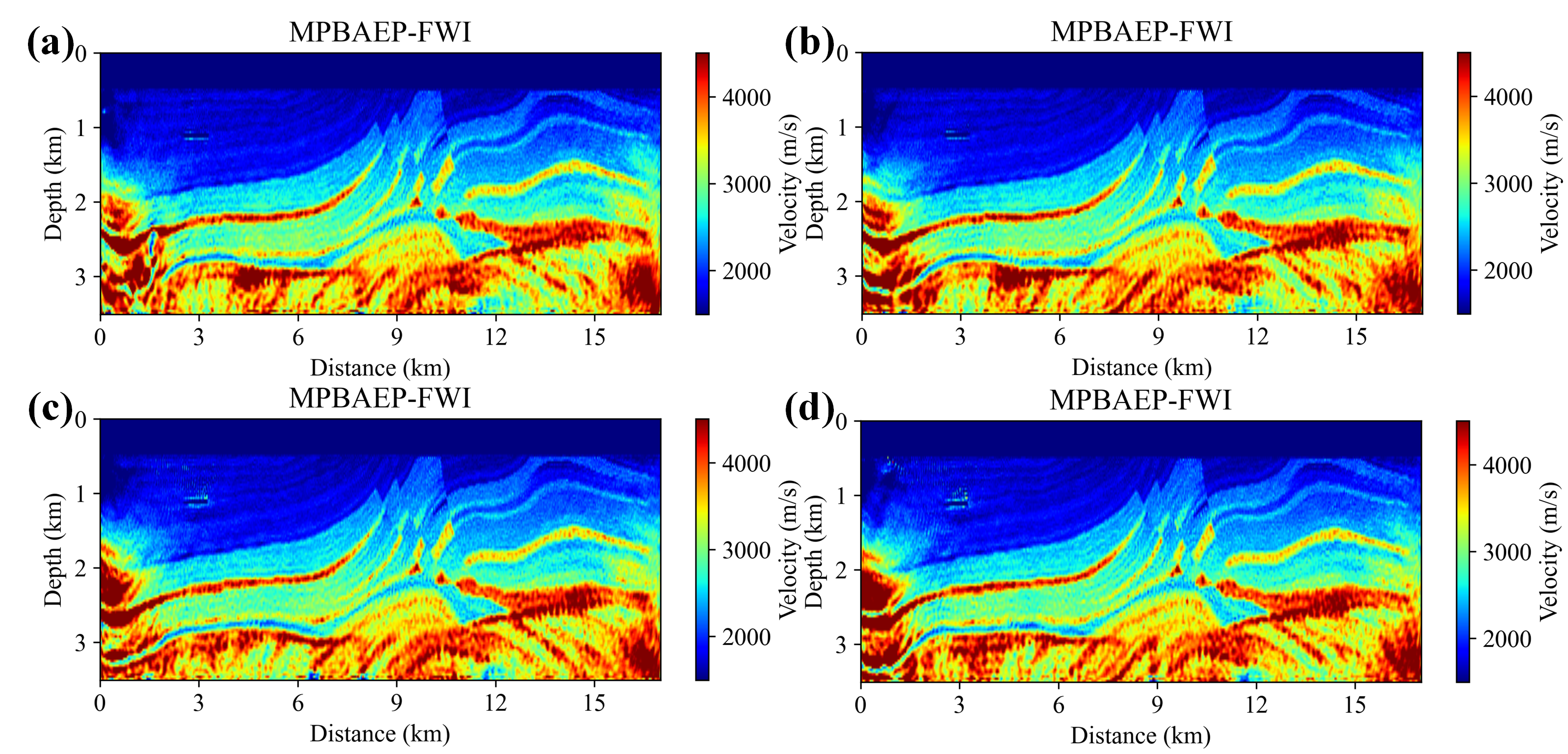}
\caption{MPBAEP-FWI inversion results using patch sizes of (a) 1×1, (b) 8×8, (c) 32×32, and (d) 128×128, respectively.}
\label{fig18}
\end{figure*}

Here, we investigate the impact of different patch sizes on MPBAEP-FWI performance. Using the Marmousi model as an example, Fig. \ref{fig18} shows the inversion results obtained with MPBAEP-FWI using shot patching under different patch sizes. The results indicate that the inverted models are nearly identical across all tested patch sizes, suggesting that, for patch sizes smaller than 128×128, the choice of patch size has negligible influence on inversion accuracy.

Table \ref{tab3} summarizes the SNR, SSIM, and RMSE metrics of MPBAEP-FWI inversion results obtained with different patch sizes. The results indicate that the inversion accuracy remains essentially invariant with respect to patch size, while the computational cost shows no significant change.

\begin{table}
\centering 
\caption{Comparison of MPBAEP-FWI inversion results for different patch sizes. The results are computed using a single NVIDIA GPU A100.}
\begin{tabular}{@{}cccccc@{}}
\hline
Patch size  & SNR (dB) & SSIM   & RMSE (km/s) & FWI iterations & Running time (s) \\ \hline
(1, 1)     & 16.9701  & 0.5888 & 0.1417    & 200            & 1031            \\
(8, 8)  & 16.8295  & 0.5457 & 0.1441    & 200            & 1041             \\
(32, 32)  & 16.4713  & 0.5300 & 0.1501    & 200            & 1039             \\
(128, 128) & 16.7436  & 0.5328 & 0.1455   & 200            & 1036             \\ 
\hline
\end{tabular}
\label{tab3}
\end{table}

\subsection{Combining MPBAEP-FWI with total variation regularization}
\begin{figure*}
\centering
\includegraphics[width=1\textwidth]{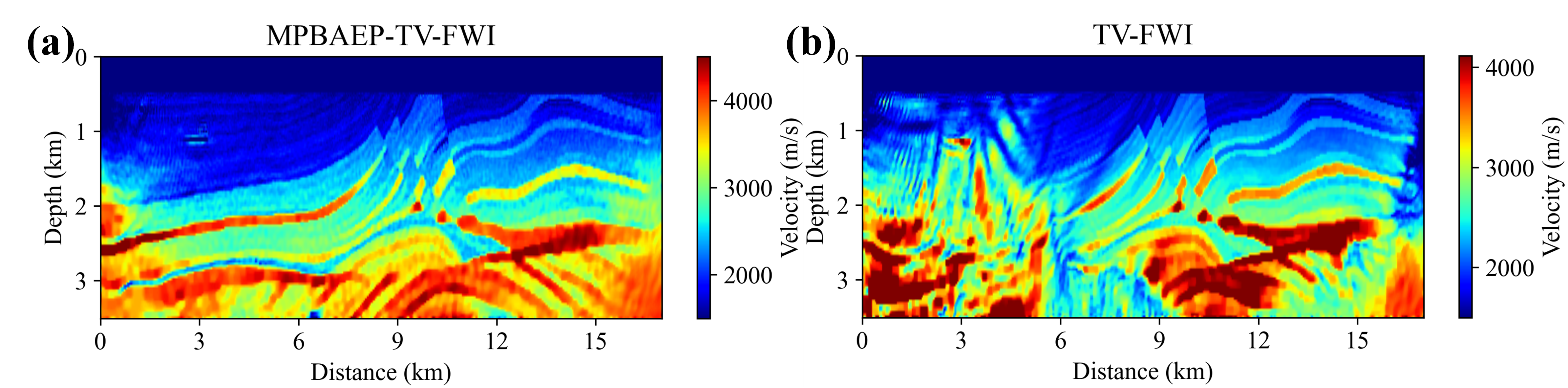}
\caption{Inversion results for the Marmousi model obtained using (a) MPBAEP-TV-FWI and (b) TV-FWI.}
\label{fig19}
\end{figure*}

As shown in Fig. \ref{fig13}(d), the presence of noise in the gradient of MPBAEP-FWI leads to noisy artifacts in the inversion result (Fig. \ref{fig10}(f)), thereby degrading inversion accuracy and resolution. To suppress this noise, we incorporate total variation (TV) regularization into the inversion framework \citep{esser2018total}. This approach is referred to as MPBAEP-TV-FWI, and the corresponding inversion result is shown in Fig. \ref{fig19}(a). Compared with Fig. \ref{fig10}(f), the noise is effectively suppressed, yielding a higher-accuracy inversion result. The SNR, SSIM, and RMSE of the MPBAEP-TV-FWI result are 17.8836, 0.6100, and 0.1276, respectively. Additionally, we combine FWI with TV regularization and refer to it as TV-FWI. The inversion result is shown in Fig. \ref{fig19}(b). Compared with the FWI result (Fig. \ref{fig10}(c)), TV-FWI improves the inversion accuracy to some extent, but it does not resolve the cycle-skipping issue.

\section{Conclusion}
We used a sequence of 2D max-pooling operations to extract an approximate envelope (MPBAE) from seismic data and formulated a corresponding objective function for full waveform inversion (FWI). Compared with the Hilbert-transform envelope (HTE), the MPBAE contains richer low-frequency components. Consequently, FWI based on the MPBAE loss (MPBAE-FWI) can further reduce the dependence on the initial velocity model compared to FWI based on the HTE loss (HTE-FWI). Convexity analysis of the objective function indicates that the MPBAE loss exhibits improved convexity compared to the HTE loss. Furthermore, by combining the Euclidean loss, the MPBAE loss, and shot patching, we develop the MPBAEP loss, which exploits the inherent energy normalization of the adjoint source associated with the Euclidean loss, thereby generating gradients with better-balanced amplitudes and improving inversion accuracy. Numerical experiments on synthetic examples and a field dataset demonstrate that MPBAE-FWI mitigates cycle skipping more effectively than HTE-FWI, thereby reducing sensitivity to the initial model. Furthermore, MPBAEP-FWI provides additional improvements in inversion accuracy compared with MPBAE-FWI.

\section{Acknowledgments}
The authors sincerely appreciate the support from KAUST and the DeepWave Consortium sponsors. They also thank the SWAG group for fostering a collaborative research environment.

\bibliography{references.bib}
\bibliographystyle{unsrt} 






\end{document}